\documentclass{aa}
\usepackage{graphicx,natbib}
\bibpunct{(}{)}{;}{a}{}{,}
\begin{document}

\title{First Polarimetry results of two candidate \\high-mass protostellar objects}
  
\author{R.L. Curran\inst{1}
 \and A. Chrysostomou\inst{1}
 \and J.L. Collett\inst{1}
 \and T. Jenness\inst{2}
 \and D.K. Aitken\inst{1}}

\offprints{\email{rcurran@star.herts.ac.uk}}

\institute{Centre for Astrophysics Research, Science and Technology Research Centre, University of Hertfordshire, College Lane, Hatfield, HERTS, AL10 9AB, UK
 \and Joint Astronomy Centre, 660 N. A`oh\={o}k\={u} Place, Hilo, HI 96720, USA} 

\date{Received / Accepted }

\abstract{

We present $850\mu$m imaging polarimetry of two high-mass star forming
regions -- W48 and S152. Within these regions we have identified two
candidate high-mass protostellar objects -- W48W and S152SE. The
submillimetre continuum emission from the candidate HMPOs is bright in
comparison to the nearby \ion{H}{ii} regions. W48W is a cold dense source,
with no radio or mid-infrared emission. S152SE has an IRAS source IRAS
22566\,+5828 in the Southwestern part of the region, which appears in
the mid-infrared $8.28\mu$m emission, but there is no radio
emission. The $850\mu$m data shows another core within the region, in
the Northeast. The polarimetry is ordered and the degree of
polarisation is high over the candidate HMPOs -- $\sim 6\%$ for W48W
and $\sim 8\%$ for S152SE.  Polarimetry results of this nature
indicate a strong, ordered magnetic field threading the candidate
HMPOs. The magnetic field direction in both S152SE and W48W is
perpendicular to the direction of elongation of the cloud which would
imply collapse along the field lines. Estimates of the magnetic field
strength are derived using the Chandrasekhar \& Fermi method. We
calculate plane of the sky field strengths of $\sim$0.7 mG for W48W
and $\sim$0.2 mG for S152SE. We discuss the drawbacks of using the
Chandrasekhar \& Fermi method with a large beam size. Mass-to-flux
ratios have been calculated and both clouds are found to be roughly
critical.

}

\titlerunning{Polarimetry of two candidate HMPOs} 
\authorrunning{R. Curran et al.}
\maketitle

\section{Introduction}

Theoretical and observational studies of low mass star formation have
lead to an evolutionary scenario based on the identification and
studies of what are referred to as Class 0-III young stellar objects
(YSOs) \citep{shu87,lada87,andre93,boss97,boss02,shirl02}. The precise
details, such as outflow mechanisms and magnetic fields are still not
definite, but understanding is improving
\citep{bach95,cern96,cabrit97,Konigl00,fiege96a,fiege96b,crutcher,mous76}. The
earliest stages of star formation are of particular interest, as this
is the stage where the highest rates of accretion occur along with the
onset of outflows. The youngest protostars - the class 0 YSOs - are
somewhat difficult to observe since they are enshrouded with dust and
so suffer large amounts of extinction.

High mass star formation is not as clearly understood. Very few
studies have been carried out on the earliest stages of high mass star
formation compared to low mass
\citep[e.g.][]{beuther02a,evans02}. This is generally because regions
of high mass star formation are located at greater distances than the
low mass star forming regions, and given that high mass star formation
takes place on shorter evolutionary timescales, coupled with the
statistical rarity of massive stars, it is difficult to locate and
observe high mass stars in the process of formation -- the so-called
high-mass protostellar objects (HMPOs). Surveys searching for HMPOs
are attempting to overcome this, in particular \citet{sridharan02} and
\citet{beuther02b}, who use a colour selected, radio-quiet IRAS
sample, and \citet{thompson} who are searching for HMPOs in the
vicinity of ultra-compact \ion{H}{ii} (UC\ion{H}{ii}) regions.

Mapping magnetic fields may yield vital information on the function of
magnetic fields in star formation. Submillimetre polarimetry is one of
the best ways to study the magnetic field. Spinning dust grains align
themselves with the magnetic field, producing polarised thermal
emission which can then be measured and used to trace the magnetic
field structure projected onto the plane of the sky.

In a bid to understand the magnetic field structures in the envelopes
of YSOs we have obtained submillimetre imaging polarimetry of a sample
of high and low mass YSOs using the JCMT in Hawaii. In this paper we
present results from two high mass regions -- W48 and S152. The two
sources were selected as part of a larger sample of high-mass star
forming regions. They were not selected specifically to find candidate
HMPOs.

\section{Observations and Data Reduction}

The observations were carried out at the James Clerk Maxwell Telescope
(JCMT) in Hawaii using the Submillimetre Common User Bolometer Array
(SCUBA) camera \citep{holland} together with the polarimetry module
\citep{jane} mounted on the entrance window of SCUBA. The data were
obtained using the jiggle mapping mode, and a 16 point jiggle map was
completed for each of the 16 different positions of the half-wave
retarder (separated by 22.5$^{\circ}$). This process was repeated 8
times for W48, giving an integration time of 2.3 hours and 9 times
(2.6 hours) for S152. The chop-throw was 150\arcsec\, with a position
angle of 0$^{\circ}$ in local coordinates for W48, and 140\arcsec\,
with a position angle of 90$^{\circ}$ in azimuthal coordinates for
S152. The chop-throw size and position angle were chosen carefully to
ensure that the telescope was not chopping onto other sources,
contaminating the data.

Data reduction was carried out using SCUBA User Reduction Facility
(SURF) \citep{jennesslight} commands. The nod of the telescope and the
flatfield corrections were carried out in the usual ways, with the
extinction correction using the $\tau_{850}$ calculated from the
$\tau_{CSO}$ \citep{archibald} (measured to be $\sim$0.08 throughout
observations of W48 and $\sim$0.07 for S152) for the time of the
observation, calculated from polynomials fitted to measurements of
$\tau_{CSO}$ taken throughout each night of observing. Noise
observations were used to identify and `switch off' any excessively
noisy bolometers so that they took no further part in the data
reduction. Sky removal of the data was carried out using bolometers
with no significant flux from the source in order to remove sky noise
from the data.

\begin{figure*}
\centering \includegraphics[width=17cm]{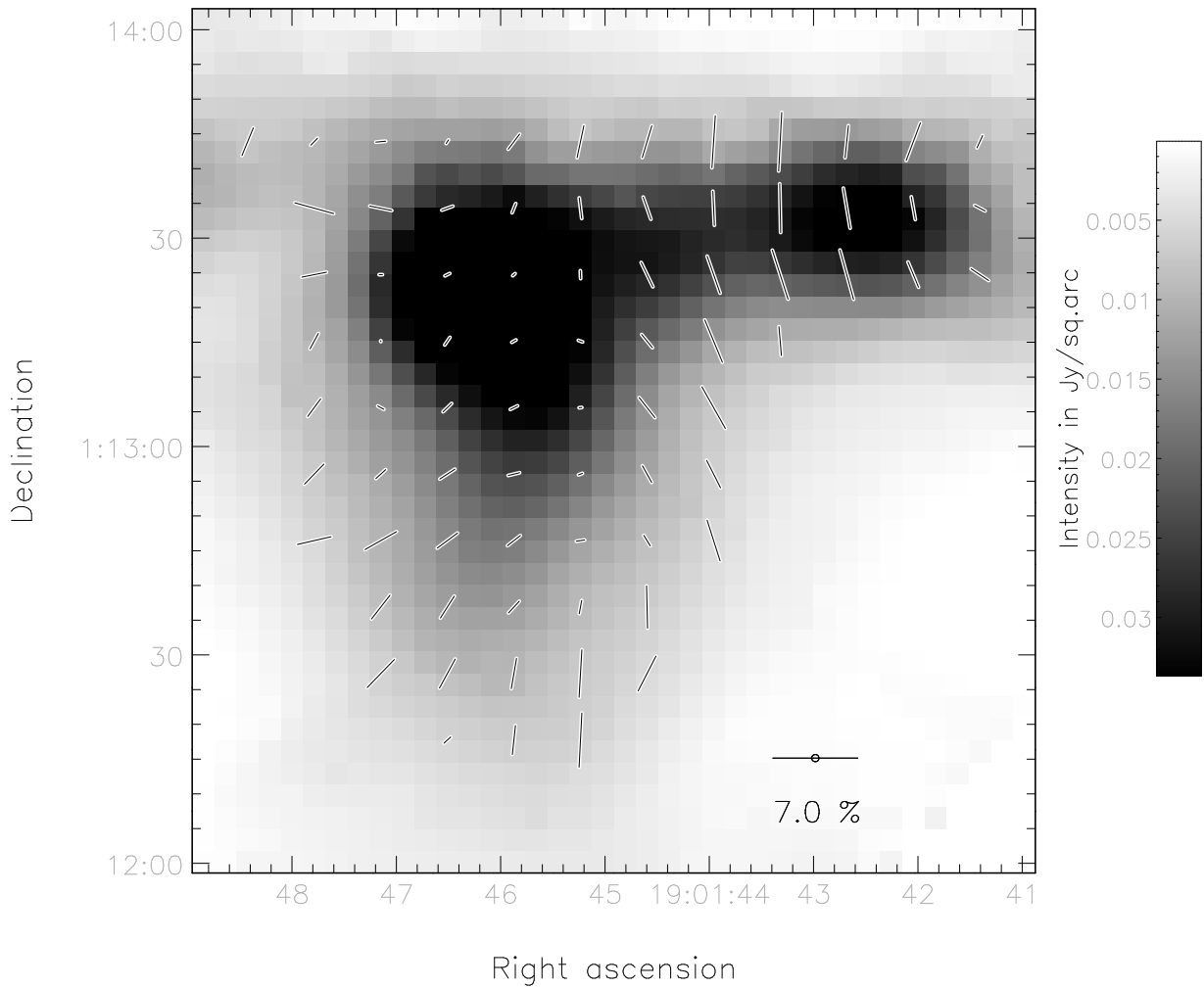} \caption{SCUBA
 polarimetry of W48. The greyscale is the 850 $\mu$m continuum
 emission, the overlaid vectors have been rotated through 90\degr to
 represent the direction of the magnetic field lines projected onto
 the plane of the sky. Epoch J2000. } \label{w48}
\end{figure*}

Once the instrumental polarisation was calculated and removed on a
bolometer-by-bolometer basis, the data were combined together to
produce final maps in the three Stoke's parameters, I, Q and U using
POLPACK software \citep{berry}. The Stoke's parameters were calculated
by fitting the following curve \citep{axon} to the data:

\begin{equation}
I'_{k} = \frac{t}{2}(I + \epsilon(Q\cos2\phi_{k} + U\sin2\phi_{k}))
\end{equation}

\noindent Where $I'_{k}$ is the expected intensity in image $k$, $t$
is the wire-grid analyser transmission factor, $\epsilon$ is the
analyser polarising efficiency factor and $\phi_{k}$ is the  effective
retarder position angle after correction for the parallactic angle for
image $k$.

These were then combined to calculate the polarised intensity, P, and
the position angles:

\begin{equation}
P=\sqrt{Q^{2}+U^{2}}
\end{equation}

\begin{equation}
\theta=\frac{1}{2}\arctan\frac{U}{Q}
\end{equation}

\noindent where $\theta$ is the position angle of the polarisation
vectors.

The uncertainties in each of these quantities are given by:

\begin{equation}
\delta P= \frac{\sqrt{\delta Q^{2}\,Q^{2}+\delta U^{2}\,U^{2}}}{P}
\end{equation}

\begin{equation}
\delta\theta =\frac{28.6^{\circ}}{s_{p}}
\end{equation}

\noindent where $s_{p}$ is the signal-to-noise in $P$.

The vectors were selected so that all have errors in polarisation of
 less than 0.5\% for W48 and and 0.75\% for S152. The error limit was
 selected to be different for the two sources as the size of the S152
 region meant that telescope offsets were required to cover the
 source, resulting in each pixel receiving less integration time
 relative to W48. The vectors are clipped on the polarisation errors
 instead of signal-to-noise as clipping on the latter would result in
 disposing of points where the polarisation is low.

Finally, flux calibration \citep{jenness} was carried out using
observations of Uranus and CRL2688 for W48 and S152 respectively. The
flux calibration factors (FCFs) on each night of observing were
calculated to be 2.02 Jy/arcsec$^{2}$/V for W48 and 1.82
Jy/arcsec$^{2}$/V for S152. These FCFs were then applied to the data.

\section{Results \& Discussion}
\subsection{Submillimetre Polarimetry}

\begin{figure*}
\centering \includegraphics[width=17cm]{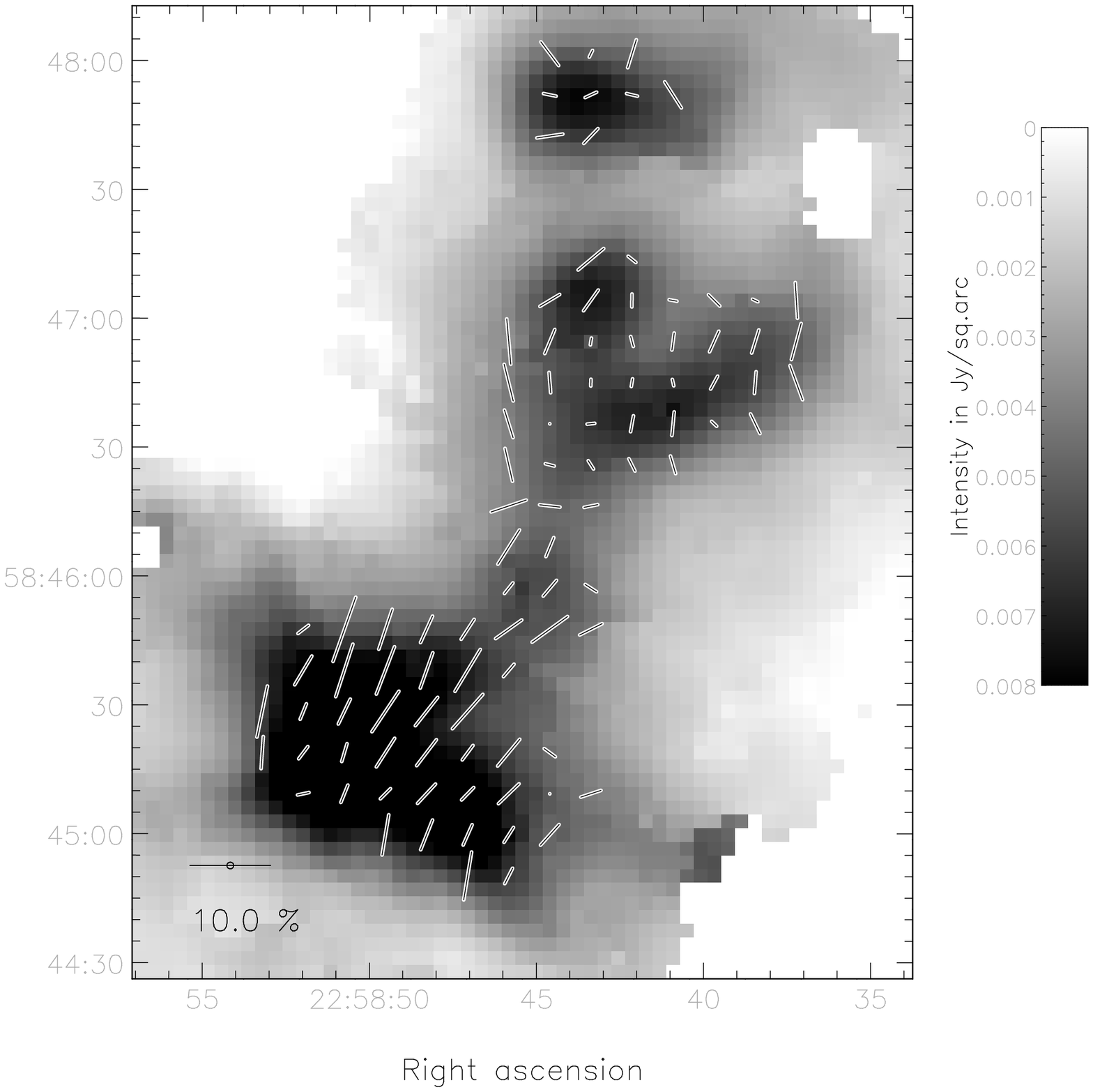} \caption{Same as
 fig. \ref{w48}, but for the region S152.}  \label{s152}
\end{figure*}

Figures \ref{w48} and \ref{s152} present the data observed with the
JCMT. The greyscale is the Stoke's I, or the intensity of the thermal
emission (850$\mu$m) from dust grains, the vectors represent the
direction of the magnetic field on the plane of the sky (the measured
{\bf E} vectors rotated through 90$^{\circ}$). The length of the
vectors indicate the degree of polarisation.

The massive star forming region W48 (Fig. \ref{w48}) is a Galactic \ion{H}{ii}
region at a distance of about 3.4 kpc \citep{vallee90} and is $\sim$
1.3$^{\circ}$ to the East of the supernova remnant W44. Numerous
infra-red sources have been detected towards W48, including IRS1 and
IRS2 \citep{zeilik}, along with H$_{2}$O and OH masers
\citep{tim,turner,genzel,hofner,caswell01}.

Fig. \ref{w48} shows that W48 has two submillimetre cores, the main,
bright core in the East, W48main, and the smaller core in the West,
W48W. The two cores appear to be connected by a ridge of dust and gas.
In the Northwest of W48main the polarimetry {\bf B}-vectors have a
North-South direction, perpendicular to the ridge connecting the two
cores. The degree of polarisation drops with increasing intensity,
from $\sim$2\% towards the edge of the source, to $\sim$0.5\% towards
the intensity peak. This has been observed in many sources, including
W51 \citep{us}, Bok globules \citep{henning} and the OMC
\citep{brenda}. Explanations for this decrease in polarisation with
increasing intensity include i) a twisting magnetic field towards the
centre of the core, resulting in a lower net polarisation percentage
-- due to cancellation effects within the beam, ii) a lower grain
alignment efficiency towards the centre of the core -- producing more
unpolarised emission, and iii) the dust particles may be more
spherical in regions of high density, which also produces more
unpolarised emission, resulting in a lower percentage polarisation.

The polarimetry across the submm source W48W and the ridge connecting
it to W48main is ordered (total variance in the position angle,
$\sigma^{2}_{\theta} \sim$ 0.05 radians$^{2}$) with a high ($\sim$6\%)
degree of polarisation. There is no evidence of a decrease in
polarisation with intensity towards the centre of W48W.

Figure \ref{s152} shows the complex region Sharpless 152
\citep{sharpless}. Located at a distance of 5kpc \citep{wouterloot93},
S152 forms part of a physical pair with the nearby region S152
\citep{pismis}. Numerous H$_{2}$O and OH masers have been studied
towards this source \citep{tim,wouterloot85,Henkel}. The S152 \ion{H}{ii}
region is located in the Northwest of the imaged region in
Fig. \ref{s152}, where the ridge splits in two. The polarimetry in
this region indicates that the magnetic field is North-South in
direction (in the plane of the sky), aligning with the North-South
ridge.

In the Southeast of the imaged region there is a submillimetre bright
core S152SE. This is the first time that this source has been observed
in the submillimetre continuum, with the S152 \ion{H}{ii} region having only
been observed in the submillimetre previously by \citet{tim}. The
polarimetry implies that the magnetic field lines are perpendicular to
the direction of elongation of the source, suggestive of collapse
along the field lines. The degree of polarisation is relatively high
($\sim$8\%) across the region S152SE. In the Southwest part of S152SE
there is an IRAS source (IRAS 22566+5828).

\subsection{Comparison with 1.4GHz and 8.28$\mu$m emission}

\begin{figure*}
\centering \includegraphics[width=17cm]{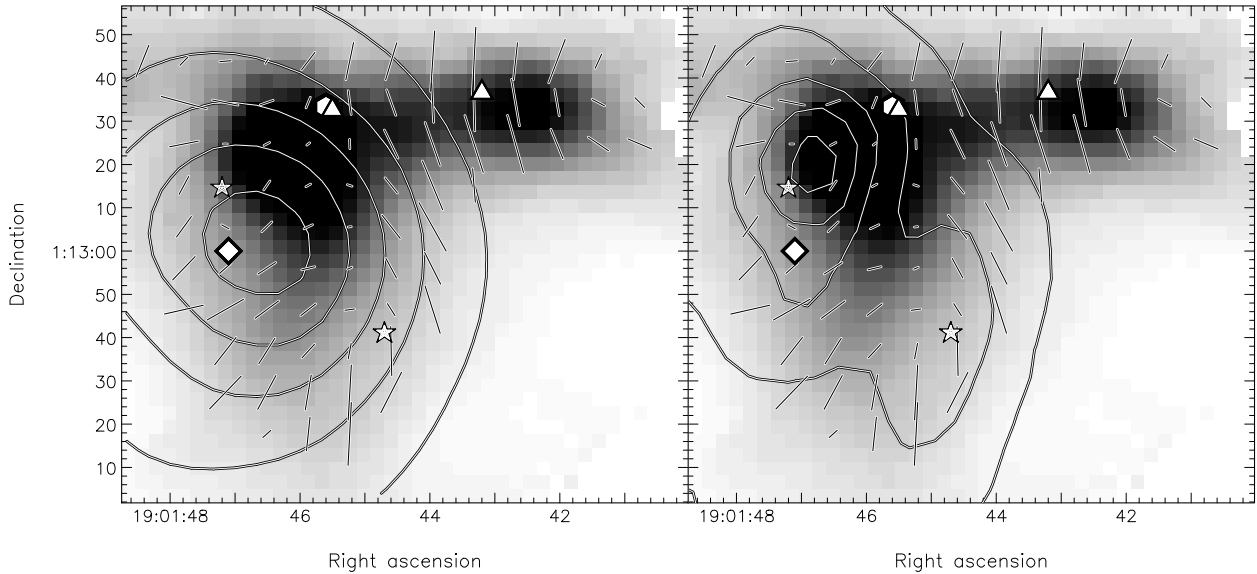} \caption{Same as
 fig \ref{w48}, but with contours of 1.4 GHz emission overlaid in the
 lefthand panel, and MSX A-band emission overlaid in the righthand
 panel. The stars represent the IRS sources \citep{zeilik}, the
 diamond marks the position of the \ion{H}{ii} region (SIMBAD database), the
 triangles represent water masers \citep{hofner}, and the hexagon
 represents OH masers \citep{caswell01}. The MSX data has a resolution
 of 20\arcsec\,, and the NVSS data has a resolution of 45\arcsec\,.}
 \label{w48cont}
\end{figure*}

\begin{figure*}
\centering \includegraphics[width=17cm]{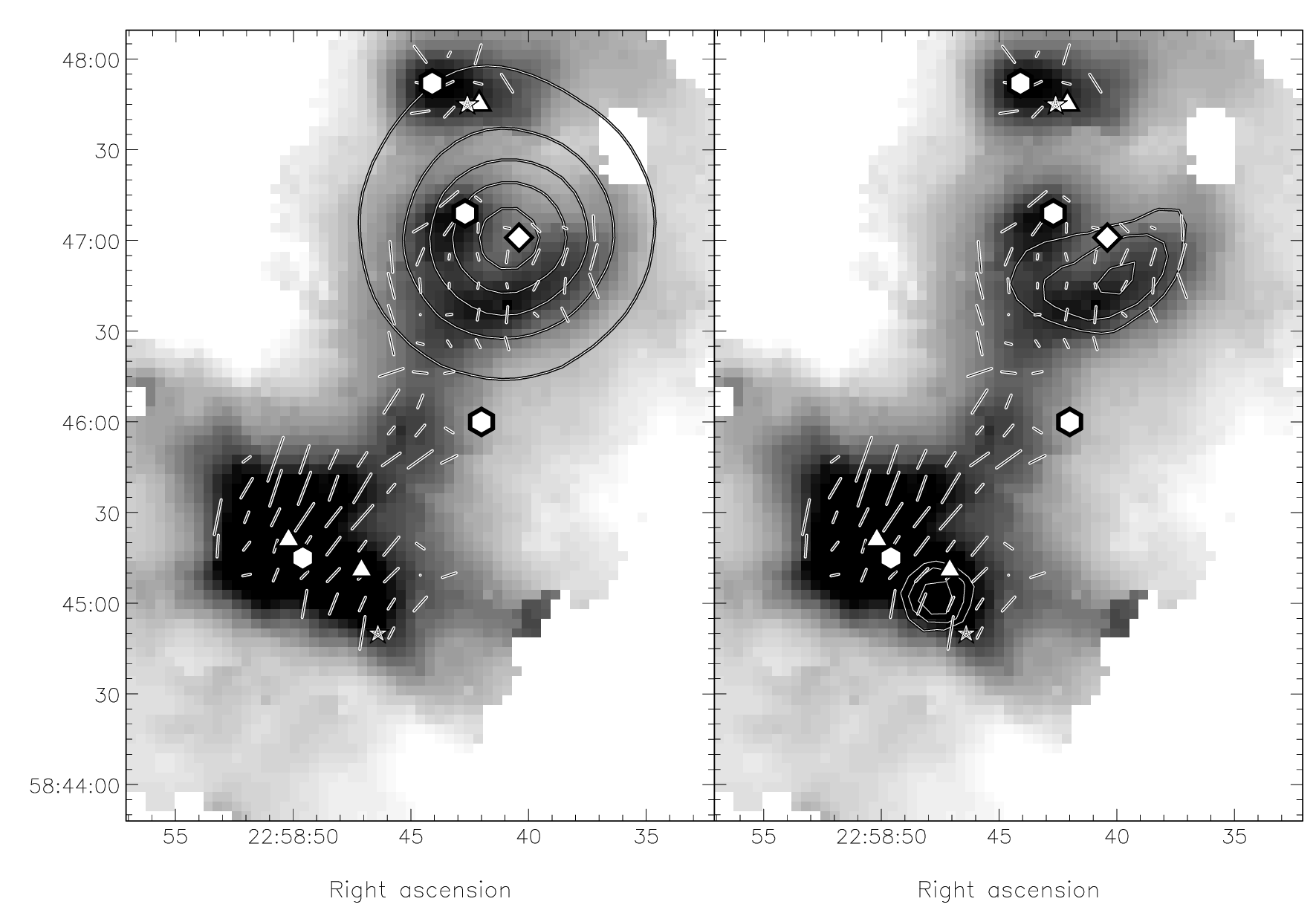} \caption{Same as
 fig \ref{w48cont}, but the stars represent the IRAS sources
 22566+5830 \& 22566+5828, the diamond marks the position of the \ion{H}{ii}
 region (SIMBAD database), the triangles represent water masers
 \citep{comoretto90}, and the hexagons represent other masers
 including OH, methanol and SiO \citep{szymczak,harju}. The MSX data
 has a resolution of 20\arcsec\,, and the NVSS data has a resolution
 of 45\arcsec\,.}  \label{s152cont}
\end{figure*}

We have obtained 1.4GHz emission maps of the W48 and S152 regions,
downloaded from the NRAO/VLA Sky Survey library (NVSS)
\citep{condon}. Also, we obtained mid-infrared data of both of these
regions at 8.28$\mu$m (A band) downloaded from the Midcourse Space
Experiment (MSX) catalogue. Figs \ref{w48cont} and \ref{s152cont} show
contour plots of the 1.4GHz and MSX A-band emission overlaid onto our
submillimetre measurements.

The left panel of Fig. \ref{w48cont} shows the W48 region imaged with
SCUBA, with contours of the 1.4GHz continuum emission. The radio
emission aligns well with the \ion{H}{ii} region to the SE of the main
submillimetre core. There is no radio emission associated with
W48W. The right panel of Fig. \ref{w48cont} shows MSX A-band (8.28
$\mu$m) mid-infrared data contoured over the SCUBA data. The infra-red
emission occurs only over the main source, indicating that the Western
source, W48W, is cold.

Figure \ref{s152cont} shows the 1.4GHz emission (left panel) and MSX
A-band data (right panel) contoured over the SCUBA data of S152. The
1.4GHz emission traces the \ion{H}{ii} region, as marked by the white
diamond. The MSX data reveal that there is infrared emission from the
dust ridges that are close to the \ion{H}{ii} region, indicating that they are
warm. There is also what appears to be a point source of warm emission
in the submillimetre bright Southeastern source S152SE. This is
probably associated with the IRAS source 22566+5828, marked by the
white star as the offset could be due to the large error of the IRAS
position, which encompasses the MSX source, or a temperature/optical
depth effect.

Comparisons of the SCUBA, MSX and NVSS 1.4GHz emission indicate that
both W48W and S152SE are cold, dense regions of gas and dust. This
suggests that they could be candidate HMPOs at an early stage of
evolution. Most current HMPO surveys \citep[e.g.][]{sridharan02} use
selection criteria which include only the more evolved HMPOs for
study. W48W and S152SE are too cold to be included in these surveys,
which are based on the IRAS Point Source Catalog, indicating that
there may be many more young HMPOs which may only be indentified
through submillimetre surveys.

\subsection{Masers}

Maser emission is common among star forming regions, specifically
H$_{2}$O, OH, methanol and ammonia maser emission. Maser emission is a
good indicator of star formation as it only occurs in regions of high
density. It has been shown that H$_{2}$O maser emission traces the
earliest evolutionary stages of massive stars
\citep{codella94,palumbo,codfel95,codella95,cfn,cfc,testi,felli}.

There are many masers within the regions observed. W48 has  H$_{2}$O
and OH masers located around the W48main submm core. Whilst no maser
emission has been detected on the source W48W, there has been a
detection of an H$_{2}$O maser close to the source, on the ridge
connecting the two cores. This indicates that the ridge and/or W48W is
the site for early star formation, supporting the idea of W48W being a
candidate HMPO. The location of the masers are indicated on
Fig. \ref{w48cont}.

Numerous masers have been detected towards S152, mainly H$_{2}$O and
SiO, with H$_{2}$O masers having been studied in S152SE. Again, only
H$_{2}$O masers have been discovered in this region indicating that
the region is possibly undergoing the earliest stages of cloud
collapse, again supporting the idea that S152SE is a candidate
HMPO. Fig \ref{s152cont} shows the location of the masers in this
region.

\subsection{Masses of Candidate HMPOs}

The mass of the clumps can be calculated via the relation:

\begin{equation}
M = \frac{g\, S_{\mathrm{\nu}}\, d^{2}}{\kappa_{\mathrm{\nu}}\,
B_{\mathrm{\nu}}\,(T_{\mathrm{dust}})}
\end{equation} 

\noindent where $g$ is the gas-to-dust ratio, $S_{\nu}$ is the flux
density of the clump, $d$ the distance to the clump, $\kappa_{\nu}$
the absorption coefficient at frequency $\nu$, and
$B_{\nu}(T_{\mathrm{dust}})$ is the Planck function at frequency $\nu$
for a blackbody of temperature of $T_{\mathrm{dust}}$.

Using a gas-to-dust ratio of 100:1 \citep{hildebrand}, and an
absorption coefficient at 850 $\mu$m of 0.15 m$^{2}$ kg$^{-1}$
estimated from \citet{ossenkopf} based on a number density of n$_{H}$
= 10$^{5}$ cm$^{-3}$, thick ice mantles and a formation timescale of
10$^{5}$ years, temperatures of $\sim$ 20K, flux densities and radii
measured from the SCUBA data listed in table \ref{sources}, along with
the distances quoted in table \ref{sources}, we calculate the masses
to be $\sim$ 1500 M$_{\sun}$ and 3500 M$_{\sun}$ respectively.

The mass of these objects are subject to the uncertainties within the
parameters used, specifically the gas-to-dust ratio, which could be as
small as 45:1 \citep{mccutcheon}. Also, the absorption coefficient at
850 $\mu$m has still not been determined precisely
\citep{hildebrand,chini}. The value adopted in this paper from
\citet{ossenkopf} agrees with the values determined by \citet{bianchi}
and \citet{visser}.

\subsection{Magnetic Field Strength}

It is important to try to establish the magnetic field strength as
well as the magnetic field morphology in these regions for a more
complete understanding of the contribution of magnetic fields to the
star formation process. Estimates of the magnetic field strength can
be gained through use of the Zeeman effect and the \citet{cf} (CF)
method. The Zeeman effect uses line splitting to calculate the
strength of the line-of-sight component of the field, while the CF
method estimates the strength of the plane of sky field based on the
dispersion of position angles of the polarisation vectors. The CF
method is applicable here and leads to estimates of 0.65 mG for W48W
and 0.2 mG for S152SE. The parameters used are shown in table
\ref{sources}.

\begin{equation}
\langle B_{\mathrm{pos}} \rangle=f \sqrt{4\pi\rho}
\frac{\sigma_{v_{\mathrm{los}}}}{\sigma_{\theta}}\;\;\;\;\;\;\mathrm{G}\label{eq:cf}
\end{equation}

\noindent where $\rho$ is the mean density (g cm$^{-3}$),
$\sigma_{v_{\mathrm{los}}}$ the line-of-sight velocity 
dispersion (cm s$^{-1}$), $\sigma_{\theta}$ is the
dispersion in polarisation position angles and is corrected for
measurement errors ($\sigma^{2}_{\theta} =
\sigma^{2}_{\mathrm{measured}} - \sigma^{2}_{\mathrm{error}}$) where
$\theta$ is in (rad), and $f$ is a correction factor found to be
$\sim$ 0.5 \citep{ostriker}.

\begin{table*}
  \caption[]{Source positions and parameters calculated from the
  observations.}  \label{sources}
	
        \begin{tabular}[t]{lllllllllll} \hline \hline Object & RA &
		Dec & Distance & Flux & Radius & Mass & Density$^{1}$
		& $\sigma_{v_{\mathrm{los}}}$ & $\sigma_{\theta}$ &
		B$_{\mathrm{pos}}$ \\ & (J2000) & (J2000) & (kpc) &
		(Jy) & (cm) & (M$_{\sun}$) & (g cm$^{-3}$) & (cm
		s$^{-1}$) & (rad) & (mG) \\ \hline W48W & 19 01 42.5 &
		+1 13 33  & 3.4 & 22 & 9 $\times$ 10$^{17}$ & 1500 & 1
		$\times$ 10$^{-18}$ & 8 $\times$ 10$^{4}$ & 0.22 &
		0.65 \\ S152SE & 22 58 50 & +58 45 22 & 5 & 22 & 2.3
		$\times$ 10$^{18}$ & 3500 & 1 $\times$ 10$^{-19}$ & 8
		$\times$ 10$^{4}$ & 0.26 & 0.2 \\ \hline \end{tabular}
		$^{1}$ Assuming spherical geometry.
\end{table*}

The CF method should be used with caution when calculating the
magnetic field strength with data of this resolution, as it can
over-estimate the field strength. The large beam size of the JCMT
means that small scale tangling of the magnetic field can occur within
a beam, so the measured vectors only represent the net magnetic field
within the beam. \citet{heitsch} used their models to test the effects
of limited resolution on the observed field structure and the CF
method of calculating the mean field strength. They conclude that
small-scale variations were completely supressed, resulting in the
impression of a more homogeneous field, and when this is the case, the
magnetic field strength is over-estimated by a factor of 2 for field
strengths consistent with molecular clouds. However, they purposefully
leave out any information on the polarisation percentage {\em p},
which would decrease where tangling of the magnetic field occurs. In
order to avoid over-estimating the magnetic field, we have only used
the candidate HMPO regions for our field strength estimates as the
field is presumed a-priori to be homogeneous, and there is no decrease
in the degree of polarisation which implies that on large-scales the
magnetic field is not tangled. We have also applied the correction
factor \citep{ostriker} of 0.5, consistent with the findings of
\citet{heitsch}.

Uncertainties in our estimates of the magnetic field strength arise
from calculating the density of the clumps, which incorporates the
errors involved in calculating the mass. Also, errors in calculating
the volume of the clump contribute as a spherical geometry has been
assumed for each clump. The velocity of the gas within the clump
introduces another error as a FWHM of $\sim$ 2 kms$^{-1}$ has been
used but it may be anywhere between 1 kms$^{-1}$ and 3 kms$^{-1}$
\citep{brand,mark}. Measurement errors are also introduced by the
angle $\theta$, although these are relatively small in comparison to
the other errors stated.

\subsection{Cloud Support}

The virial theorem provides a gross method of analysing the stability
of the clumps. We will simply estimate the values of the various
energy terms to see how much support is provided by bulk motions and
magnetic pressure against collapse.  Applying our magnetic field
strengths to the Virial Theorem:

\begin{equation}
\left|\frac{-3 G M^{2}} {5R}\right| > 2 \times 0.27 M \Delta v^{2} +
0.1 B^{2} R^{3}\label{eq:vir}
\end{equation}

\noindent \citep{mckee93} where $G$ is the gravitational constant, $M$
is the mass of the clump, $R$ the radius, $\Delta v$ the turbulent
velocity of the gas (FWHM) and B the magnetic field strength in Gauss.

Substituting values from table \ref{sources} into Eq. \ref{eq:vir}, it
can be shown that the gravitational energy is of the order of the sum
of the kinetic and magnetic energy terms, indicating the clouds are
gravitationally bound or contracting.

There are currently two models of star formation; star formation
driven by ambipolar diffusion and that driven by turbulence.  The
ambipolar diffusion model of star formation \citep{mouschovias} is
based on subcritical clouds that at first contract via ambipolar
diffusion. When the central parts become supercritical, the magnetic
field cannot support the core and it collapses dynamically. The
turbulence model of star formation states that turbulence provides
support against cloud collapse, as the magnetic field is too weak to
provide support (the core is supercritical).

To determine the importance of the magnetic field as a support
mechanism, the mass to flux ratio can be used.

\begin{equation}
\lambda =
\frac{\left(\frac{M}{\Phi}\right)_{\mathrm{actual}}}{\left(\frac{M}{\Phi}\right)_{\mathrm{critical}}}
\end{equation}

\noindent where we take the critical value to be:

\begin{equation}
\left(\frac{M}{\Phi}\right)_{\mathrm{critical}} \approx
\frac{0.4}{\sqrt{G}}\label{eq:crit}
\end{equation} 

\noindent \citep{mestel}. If the ratio is lower than the critical
value ($\lambda <$ 1), the cloud is said to be subcritical, and the
magnetic field can support the cloud perpendicular to the field
direction irrespective of external pressure. We calculate mass-to-flux
ratios, in units of the critical ratio, of $\lambda \sim$ 1.2 for
W48W, using a magnetic field strength of 0.65 mG, and $\lambda \sim$
1.3 for S152SE using a magnetic field strength of 0.2 mG. This implies
that both W48W and S152SE are close to critical.

The value of the numerical coefficient in Eq. \ref{eq:crit} is
model-dependent, with Eq. \ref{eq:crit} representing the model for a
spherical cloud inclusive of surface terms. \citet{mestel} also
calculates a numerical coefficient of 0.2 for an extreme spheroidal
model of eccentricity 1, whereas \citet{nakanonakamura} use a value of
0.4 for a disk-like model (ignoring surface terms), and
\citet{crutcher03} uses a value of 0.2 for a spherical model (ignoring
surface terms). This leads to ranges in $\lambda$ of $1.2 < \lambda <
2.3$ for W48W and $1.3 < \lambda < 2.7$ for S152SE. These calculations
of $\lambda$ are based on the plane of the sky magnetic field
strength, B$_{\mathrm{pos}}$. If the magnetic field does not lie
entirely in the plane of the sky in these regions, the actual magnetic
field strength will be greater than that calculated, and therefore the
value of $\lambda$ will decrease.

Previous studies of magnetic fields \citep[see][for a
review]{crutcher03} have all indicated critical/supercritical star
forming cores, with the only subcritical results being for the diffuse
ISM and HI clouds. \citet{crutcher03a} studied prestellar cores, and
found that after corrections for geometry applied statistically, they
were approximately critical. These are at roughly the same
evolutionary stage as HMPOs, but are the low-mass equivalent. Our
results are consistent with W48W and S152SE being very early stage
HMPOs, and are found to be close to critical.

\section{Conclusions}

We present the first submillimetre polarimetry data taken towards
HMPOs, found in the vicinity of the W48 and S152 \ion{H}{ii} regions. The
polarimetry across W48 is complicated. There is a drop in the
polarisation percentage with increasing intensity which may be
explained by twisting of the magnetic field towards the centre of the
core, or by relatively increased emission from unaligned dust
grains. Polarimetry of S152 indicates that the plane of sky magnetic
field has a North-South direction across the \ion{H}{ii} region.

Comparisons of our submillimetre continuum data with 8.28$\mu$m and
1.4GHz emission indicate that W48W and S152SE are candidate HMPOs, at
an early stage of evolution. The presence of water masers in these
areas supports this hypothesis. Both candidate HMPOs exhibit high
degrees of polarisation (6--8\%) and CF calculations provide estimates
of the plane of sky magnetic field strength of $\sim$ 0.65mG for W48W and
$\sim$ 0.2mG for S152SE.

The polarisation vectors are well aligned in both regions, indicating
the magnetic field lines are perpendicular to the direction of
elongation and imply collapse along field lines. This polarimetry
pattern has been predicted using models of subcritical cores
collapsing under the effect of gravity and ambipolar drift
\citep[see][]{padoan}, although our estimates of $\lambda$ show that
these clouds are close to critical and as such we are unable to rule
out turbulence driven star formation models.

\begin{acknowledgements}
The JCMT is operated by the Joint Astronomy Centre in Hilo, Hawaii on
behalf of the parent organizations Particle Physics and Astronomy
Research Council in the United Kingdom, the National Research Council
of Canada and The Netherlands Organization for Scientific
Research. This research made use of data products from the
Midcourse Space Experiment. Processing of the data was funded by the
Ballistic Missile Defense Organization with additional support from
NASA Office of Space Science.  This research has also made use of the
NASA/IPAC Infrared Science Archive, which is operated by the Jet
Propulsion Laboratory, California Institute of Technology, under
contract with the National Aeronautics and Space Administration.  The
authors acknowledge the data analysis facilities provided by the
Starlink Project which is run by CCLRC on behalf of PPARC.

The authors would like to thank our referee, Richard Crutcher, for
careful reading of the original manuscript and useful comments.

\end{acknowledgements}

\end{document}